\documentclass[10pt,conference]{IEEEtran}
\usepackage{epsfig,setspace,amsmath,epsf,amssymb,bm,theorem,cite,graphicx, epstopdf,algorithm,algpseudocode,float,color,mathtools}
\usepackage[table,xcdraw]{xcolor}

\newtheorem{theorem}{Theorem}

\newtheorem{remark}{Remark}
\newtheorem{lemma}{Lemma}

\IEEEoverridecommandlockouts
\allowdisplaybreaks

\begin{document}

\title{Private Read Update Write (PRUW) With Heterogeneous Databases}

\author{Sajani Vithana \qquad Sennur Ulukus\\
\normalsize Department of Electrical and Computer Engineering\\
\normalsize University of Maryland, College Park, MD 20742\\
\normalsize  \emph{spallego@umd.edu} \qquad \emph{ulukus@umd.edu}}

\maketitle

\begin{abstract}
We investigate the problem of private read update write (PRUW) with heterogeneous storage constrained databases in federated submodel learning (FSL). In FSL a machine learning (ML) model is divided into multiple submodels based on different types of data used to train it. A given user downloads, updates and uploads the updates back to a single submodel of interest, based on the type of user's local data. With PRUW, the process of reading (downloading) and writing (uploading) is carried out such that information theoretic privacy of the updating submodel index and the values of updates is guaranteed. We consider the practical scenario where the submodels are stored in databases with arbitrary (heterogeneous) storage constraints, and provide a PRUW scheme with a storage mechanism that utilizes submodel partitioning and encoding to minimize the communication cost.
\end{abstract}

\section{Introduction}

In private read update write (PRUW) \cite{pruw_jpurnal, dropout, pruw, sparse, rd, ourICC, segmentation, rate_privacy_storage}, the users read from and write to selected sections of a data storage without revealing what is read or written. We use PRUW in federated submodel learning (FSL)\cite{secureFSL, paper1, dropout, pruw_jpurnal, billion, recent} to guarantee user privacy in the learning process. FSL is a method developed to reduce the communication cost in federated learning (FL), in which the central FL model is divided into multiple submodels based on different types of data used to train it. In FSL, a given user only downloads the submodel relevant to the user's local data, updates it, and writes (uploads) the updates back to the submodel. The updating submodel index and the uploaded values of updates leak information about the type of data the user has. With PRUW, the read-write process in FSL is carried out while guaranteeing information theoretic privacy of the updating submodel index and the values of updates. In PRUW, the submodels are stored in multiple non-colluding databases, similar to private information retrieval (PIR)\cite{original, sun2017capacity, chaoTian, Kumar_PIRarbCoded, SPIR, PSI, MMPIR, YamamotoPIR, VardyConf2015, coded, RobustPIR_Razane, MultiroundPIR, arbmsgPIR, WangSkoglund, colluding, codedcolluded, codedcolludedJafar, tandon2017capacity, wei2017fundamental, leaky, ravi-leaky, sideinfo, securePIRcapacity, evesdroppers, byzantine, semanticPIR, singleDB,XSTPIR, CSA, smallfields, SecureStorage, MAC_PIR_ITW2018}.  

In this paper, we consider the practical scenario where the multiple non-colluding databases have arbitrary storage constraints. PRUW with homogeneous storage constraints is studied in \cite{pruw}. PIR with homogeneous and heterogeneous storage constrained databases is studied in \cite{heteroPIR, utah, efficient_storage_ITW2019, PIR_decentralized,Tandon2018PIRFS,StorageConstrainedPIR,StorageConstrainedPIR_Wei}, which introduce optimum partitioning strategies to fit the messages into the storage constrained databases in such a way that the download cost is minimized. In this work, we extend the ideas used in PIR with storage constrained databases to PRUW. The two main methods used in the PIR literature to handle storage constrained databases are: 1) coded storage, where the model parameters are encoded with a specific $(K,R)$ MDS code, and 2) divided storage, where the model is partitioned, and different partitions are replicated in different subsets of databases. We first observe that there is an advantage of using $(K,R)$ MDS coded storage with higher values of $K$ in PRUW with heterogeneous databases, and propose a PRUW scheme and a storage mechanism for FSL that is based on both submodel partitioning and MDS coding to efficiently utilize the available space in all databases with the goal of minimizing the total communication cost. 

\begin{figure}[t]
    \centering
    \includegraphics[width=0.8\linewidth]{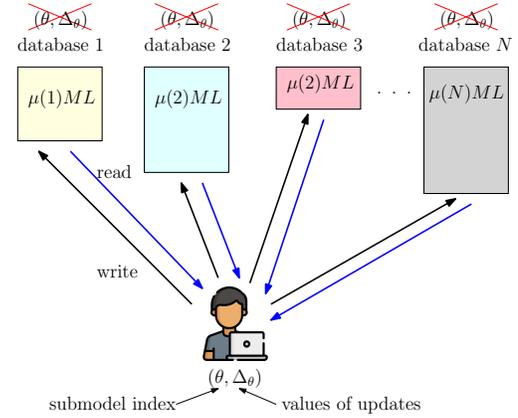}
    \caption{System model.}
    \label{model}
    \vspace*{-0.5cm}
\end{figure}

\section{Problem Formulation}

We consider a FSL setting with $M$ independent submodels, each containing $L$ parameters, taking values from a finite field $\mathbb{F}_q$, stored in a system of $N$ non-colluding databases. Database $n$ has a storage capacity of $\mu(n) ML$ symbols, where $\mu(n)\leq1$ for $n\in\{1,\dotsc,N\}$; see Fig.~\ref{model}. That is,
\begin{align}
    H(S_n^{[t]})\leq \mu(n) ML, \quad n\in\{1,\dotsc,N\},\quad t\in\mathbb{N},    
\end{align}
where $S_n^{[t]}$ is the content stored in database $n$ at time $t$.  At a given time $t$, a user downloads a required submodel without revealing its index to any of the databases. This is the reading phase of the PRUW process. The user then updates the submodel locally, and uploads the updates back to the same submodel in all databases without revealing the values of updates or the submodel index. This is the writing phase. 

\emph{Privacy of the submodel index:} No information on the updating submodel index at time $t$, $\theta^{[t]}$, is allowed to leak to any of the databases, i.e., for each $n$, $n\in\{1,\dotsc,N\}$,
\begin{align}
    I(\theta^{[t]};Q_n^{[t]},U_n^{[t]}|Q_n^{[1:t-1]},S_n^{[0:t-1]},U_n^{[1:t-1]})=0, 
\end{align}
where $Q_n$ and $U_n$ are the queries and updates sent by the user to database $n$ at time instances denoted in square brackets.

\emph{Privacy of the values of updates:} No information on the values of updates is allowed to leak to any of the databases,
\begin{align}    
I(\Delta_{\theta}^{[t]};U_n^{[t]}|Q_n^{[1:t]},S_n^{[0:t\!-\!1]},U_n^{[1:t\!-\!1]})\!=\!0,\! \quad n\!\in\!\{\!1,\dotsc,N\!\}
\end{align}
where $\Delta_{\theta}^{[t]}$ is the update generated by the user at time $t$.

\emph{Security of submodels:} No information on the values of submodels is allowed to leak to any of the databases, i.e., 
\begin{align}
I(W_{1:M}^{[t]};S_n^{[t]})=0, \quad n\in\{1,\dotsc,N\}
\end{align}
where $W_k^{[t]}$ is the $k$th submodel at time $t$. 

\emph{Correctness in the reading phase:} The user should correctly download the required submodel in the reading phase, i.e., 
\begin{align}
    H(W_{\theta}^{[t-1]}|Q_{1:N}^{[t]},A_{1:N}^{[t]})=0
\end{align}
where $A_n^{[t]}$ is the answer received from database $n$ at time $t$.

\emph{Correctness in the writing phase:} The submodels in all databases must be correctly updated as,
\begin{align}
    W_m^{[t]}=\begin{cases}
        W_m^{[t-1]}+\Delta_m^{[t]}, & \text{if 
 $m=\theta^{[t]}$}\\
        W_m^{[t-1]}, & \text{if $m\neq\theta^{[t]}$}.
    \end{cases}
\end{align}

The reading and writing costs are defined as $C_R=\frac{\mathcal{D}}{L}$ and $C_W=\frac{\mathcal{U}}{L}$, where $\mathcal{D}$ and $\mathcal{U}$ are the total number of symbols downloaded and uploaded in the reading and writing phases, respectively. The total cost $C_T$ is the sum of the reading and writing costs, i.e., $C_T=C_R+C_W$. In this paper, we use the notation $C_T(a,b)$ to indicate the total cost when the submodels are encoded with an $(a,b)$ MDS code.

\section{Main Result}

\begin{theorem}\label{thm1}
    For a given PRUW setting with $N$ non-colluding databases having arbitrary storage constraints $\mu(n)$, $n\in\{1,\dotsc,N\}$, let $k=\frac{1}{\max_{n}\mu(n)}$, $p=\sum_{n=1}^N \mu(n)$, $r=kp$ and $s=\lfloor k\rfloor p$. Then, there exist a storage mechanism and a PRUW scheme that completely fills the $N$ databases and achieves a total communication cost of $\mathcal{C}=\min\{C_1,C_2\}$, where 
    \begin{align}\label{c1}
        C_1&=(\lceil s\rceil-s)C_T(\lfloor k\rfloor,\lfloor s\rfloor)+(s-\lfloor s\rfloor)C_T(\lfloor k\rfloor,\lceil s\rceil)\\
        C_2&=\alpha\beta C_T(\lfloor k\rfloor, \lfloor r\rfloor)+\alpha(1-\beta) C_T(\lfloor k\rfloor,\lceil r\rceil)\nonumber\\
        &\quad+(1-\alpha)\delta C_T(\lceil k\rceil,\lfloor r\rfloor)+(1-\alpha)(1-\delta) C_T(\lceil k\rceil,\lceil r\rceil)\label{c2}
    \end{align}
where, 
\begin{align}
    \alpha&=\begin{cases}
        \frac{\lfloor k\rfloor(p\lceil k\rceil-\lceil r\rceil)}{\lceil k\rceil\lfloor r\rfloor-\lfloor k\rfloor\lceil r\rceil}, & \text{if $r-\lfloor r\rfloor>k-\lfloor k\rfloor$, $s\leq\lfloor r\rfloor$}\\
        \frac{\lfloor k\rfloor}{k}(\lceil k\rceil-k), & \text{else}
    \end{cases}\\
    \beta&=\begin{cases}
        \frac{\lceil r\rceil-r}{\lceil k\rceil-k}, & \text{if  $r-\lfloor r\rfloor>k-\lfloor k\rfloor$, $s>\lfloor r\rfloor$}\\
        1, & \text{else}
    \end{cases}\\
    \delta&=\begin{cases}
        1-\frac{r-\lfloor r\rfloor}{k-\lfloor k\rfloor}, & \text{if  $r-\lfloor r\rfloor\leq k-\lfloor k\rfloor$}\\
        0, & \text{else},
    \end{cases}
\end{align}
 if $\lfloor r\rfloor-\lfloor k\rfloor$ is odd, and 
\begin{align}
    \alpha&=\begin{cases}
        \frac{\lfloor k\rfloor}{k}(\lceil k\rceil-k), & \text{if $r-\lfloor r\rfloor<\lceil k\rceil-k$}\\
        \frac{\lfloor k\rfloor(p\lceil k\rceil-\lfloor r\rfloor)}{\lceil k\rceil\lceil r\rceil-\lfloor k\rfloor\lfloor r\rfloor}, & \text{else}
    \end{cases}\\
    \beta&=\begin{cases}
        1-\frac{r-\lfloor r\rfloor}{\lceil k\rceil-k}, & \text{if  $r-\lfloor r\rfloor<\lceil k\rceil-k$}\\
        0, & \text{else}
    \end{cases}\\
    \delta&=1
\end{align}
when $\lfloor r\rfloor-\lfloor k\rfloor$ is even, with the function $C_T(a,b)$ defined as, 
\begin{align}\label{totalcost}
        C_T(a,b)=\begin{cases}
            \frac{4b}{b-a-1}, & \text{if $b-a$ is odd}\\
            \frac{4b-2}{b-a-2}, & \text{if $b-a$ is even}.
        \end{cases}
    \end{align}
\end{theorem}

\begin{figure}[t]
    \centering
    \includegraphics[width=0.8\linewidth]{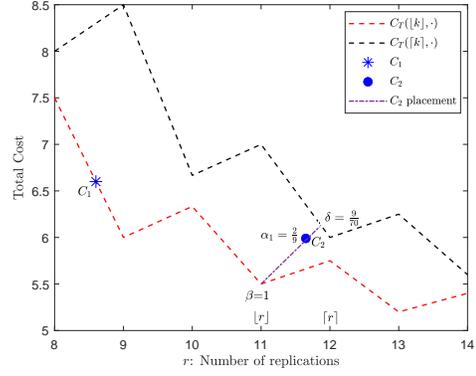}
    \caption{Example setting with $k=2.7$ and $p=4.3$.}
    \label{example}
    \vspace*{-0.3cm}
\end{figure}

\begin{remark}\label{rem1}
As an illustration of the main result, consider an example with $\max_n \mu(n)=0.37$ and $p=\sum_{n=1}^N \mu(n)=4.3$. Then, $k=2.7$, $r=11.61$ and $s=8.6$, which results in $C_1=6.6$ as shown by the blue star in Fig.~\ref{example}. Since $\lfloor r\rfloor-\lfloor k\rfloor=9$ is odd, $s=8.6<11=\lfloor r\rfloor$ and $r-\lfloor r\rfloor=0.61<0.7=k-\lfloor k\rfloor$, for the calculation of $C_2$, $\alpha$, $\beta$ and $\delta$ are given by $\alpha=\frac{\lfloor k\rfloor}{k}(\lceil k\rceil-k)=\frac{2}{9}$, $\beta=1$ and $\delta=1-\frac{r-\lfloor r\rfloor}{k-\lfloor k\rfloor}=\frac{9}{70}$, respectively. Hence, $C_2=5.99$ (blue dot in Fig.~\ref{example}). Note that $C_1$ is obtained by storing $\lceil s\rceil-s$ and $s-\lfloor s\rfloor$ fractions of parameters of all submodels using $(\lfloor k\rfloor, \lfloor s\rfloor)$ and $(\lfloor k\rfloor, \lceil s\rceil)$ MDS codes, respectively. Similarly, $C_2$ is obtained by storing $\alpha$, $(1-\alpha)\delta$ and $(1-\alpha)(1-\delta_1)$ fractions of all submodels using $(\lfloor k\rfloor,\lfloor r\rfloor)$, $(\lceil k\rceil,\lfloor r\rfloor)$ and $(\lceil k\rceil,\lceil r\rceil)$ MDS codes, respectively, as shown in Fig.~\ref{example}. The minimum achievable cost for this example is $\min\{C_1,C_2\}=C_2$.
\end{remark}

\begin{remark}
    The total cost decreases with the number of replications $b$ and increases with the coding parameter $a$ as shown in \eqref{totalcost}. However, when the number of replications ($r$ or $s$) is chosen as a linear function of the coding parameter $k$, i.e., $r=kp$, $s=\lfloor k\rfloor p$, the total cost decreases with the coding parameter; see Section~\ref{encode}. Therefore, $C_2$ in the main result uses the maximum $k$ by considering both $\lfloor k\rfloor$ and $\lceil k\rceil$ when $k\notin\mathbb{Z}^+$. However, based on the fluctuating structure of the total cost in \eqref{totalcost} (dotted lines in Fig.~\ref{example}) for special cases of $k$ and $p$, a lower total cost can be achieved by only considering $\lfloor k\rfloor$. These cases are handled by $C_1$. 
\end{remark}

\begin{remark}
    For a given PRUW setting with arbitrary storage constraints $\{\mu(n)\}_{n=1}^N$, let the total available storage be $p=\sum_{n=1}^N \mu(n)$. Then, there exists an uncoded PRUW scheme (the proposed scheme in Section~\ref{schm} with $k=1$ and $r=p$) that replicates $\lceil p\rceil-p$ and $p-\lfloor p\rfloor$ fractions of uncoded submodel parameters in $\lfloor p\rfloor$ and $\lceil p\rceil$ databases, respectively, achieving a total cost of $(\!\lceil p\rceil\!-\!p)C_T(1,\!\lfloor p\rfloor)\!+\!(p\!-\!\lfloor p\rfloor)C_T(1,\!\lceil p\rceil)$. Note that this total cost only depends on the sum of all arbitrary storage constraints (not on individual constraints), which is consistent with the corresponding result of PIR with uncoded heterogeneous storage constraints \cite{heteroPIR}. However, in PRUW, for a given $p$, the total cost is minimized when $k\!=\!\frac{1}{\max_n\mu(n)}$ is maximized, i.e., when all storage constraints are equal. 
\end{remark}

\section{Proposed Scheme}\label{schm}

The proposed scheme for FSL with heterogeneous storage constrained databases comprises two main components, namely, the storage mechanism and the PRUW scheme. In this work, we use the PRUW scheme presented in \cite{pruw} and \cite{dropout}, and propose a storage mechanism to perform FSL on any given setting with arbitrary storage constraints. In this section, we first state the PRUW scheme briefly for completeness and then present the proposed storage mechanism.

\subsection{PRUW Scheme}

The scheme is defined for a single subpacket of all submodels (collection of $y$ coded bits), which is then applied on all subpackets identically.

\emph{Storage:} The contents of a single subpacket in database $n$, $n\in\{1,\dotsc,R\}$, at time $t$ is given by,
\begin{align}
S_n^{[t]}=\begin{bmatrix}
    \sum_{i=1}^K\frac{1}{f_{1,i}-\alpha_n}\begin{bmatrix}
    W_{1,1}^{[i]}\\\vdots\\W_{M,1}^{[i]}
    \end{bmatrix}+\sum_{j=0}^y\alpha_n^j Z_{1,j}\\
    \vdots\\
    \sum_{i=1}^K\frac{1}{f_{y,i}-\alpha_n}\begin{bmatrix}
    W_{1,y}^{[i]}\\\vdots\\W_{M,y}^{[i]}
    \end{bmatrix}+\sum_{j=0}^y\alpha_n^j Z_{y,j}
    \end{bmatrix},\label{storage}
\end{align}
where $W_{m,j}^{[i]}$ is the $i$th parameter of the $j$th coded symbol of submodel $m$, $Z_{i,j}$ are random noise vectors of size $M\times1$ and $f_{i,j}$s and $\alpha_n$s are globally known distinct constants from $\mathbb{F}_q$. Note that the $j$th coded data symbol (without the noise) is stored in terms of a $(K,R)$ MDS code. 

\emph{Reading phase:} In the reading phase, the user sends queries to databases to download the required submodel $W_{\theta}$. The queries sent to database $n$, $n\in\{1,\dotsc,R\}$, to download the $Ky$ parameters of each subpacket of $W_{\theta}$ is given by,
\begin{align}           
    Q_{n,\ell}=\begin{bmatrix}
    \frac{\prod_{i=1,i\neq \ell}^{K}(f_{1,i}-\alpha_n)}{\prod_{i=1,i\neq \ell}^{k}(f_{1,i}-f_{1,\ell})}e_M(\theta)+\prod_{i=1}^K (f_{1,i}-\alpha_n)\tilde{Z}_{1,\ell}\\
    \vdots\\
    \frac{\prod_{i=1,i\neq \ell}^{K}(f_{y,i}-\alpha_n)}{\prod_{i=1,i\neq \ell}^{k}(f_{y,i}-f_{y,\ell})}e_M(\theta)+\prod_{i=1}^K (f_{y,i}-\alpha_n)\Tilde{Z}_{y,\ell}
\end{bmatrix}\label{query}
\end{align}
for $\ell\in\{1,\dotsc,K\}$ where $e_M(\theta)$ is the all zeros vector of size $M\times1$ with a 1 at the $\theta$th position and $\Tilde{Z}_{j,\ell}$ are random noise vectors of size $M\times1$. Then, database $n$ generates the corresponding answers given by,
\begin{align}
    A_{n,\ell}&=S_n^T Q_{n,\ell}=\sum_{j=1}^y\frac{1}{f_{j,\ell}-\alpha_n}W_{\theta,j}^{[\ell]}+P_{\alpha_n}(K+y), \label{last_ans}
\end{align}
for each $\ell\in\{1,\dotsc,K\}$, where $P_{\alpha_n}(h)$ is a polynomial in $\alpha_n$ of degree $h$. Lemma 1 in \cite{pruw_jpurnal} is used to obtain \eqref{last_ans}. The user obtains the $Ky$ parameters of each subpacket of $W_{\theta}$ by downloading the answers in \eqref{last_ans} from $R'=2y+K+1$ databases, since there are $y$ parameters of $W_{\theta}$ and $K+y+1$ polynomial coefficients to be solved for, in each set of answers for a fixed $\ell$. Since the total number of databases considered is $R$, $R'=R$ if $R-K$ is odd and $R'=R-1$ if $R-K$ is even, since $y\in\mathbb{Z}^+$. Therefore, the subpacketization $y$ is given by $y=\frac{R'-K-1}{2}$, which results in a reading cost of,
\begin{align}
    C_R=\frac{R'\times K}{y\times K}=\frac{2R'}{R'-K-1}.
\end{align}

\emph{Writing phase:} In the writing phase, the user sends $K$ symbols to each of the $R$ databases. The $\ell$th symbol (out of the $K$ symbols uploaded) is the combined update that consists of the updates of 
$W_{\theta}$ corresponding to the $\ell$th parameter of each of the $y$ coded parameters in each subpacket. The $K$ combined updates sent to database $n$, $n\in\{1,\dotsc,R\}$, are given by,
\begin{align}\label{update}
    U_{n,\ell}=\sum_{j=1}^y\prod_{i=1,i\neq j}^y(f_{i,\ell}-\alpha_n)\Tilde{\Delta}_{\theta,j}^{[\ell]}+\prod_{i=1}^y(f_{i,\ell}-\alpha_n)\hat{z}_{\ell},
\end{align}
for $\ell\in\{1,\dotsc,K\}$ where $\Tilde{\Delta}_{\theta,j}^{[\ell]}=\frac{\prod_{i=1,i\neq \ell}^K (f_{j,i}-f_{j,\ell})}{\prod_{i=1,i\neq j}^y (f_{i,\ell}-f_{j,\ell})}\Delta_{\theta,j}^{[\ell]}$ for $j\in\{1,\dotsc,y\}$, with $\Delta_{\theta,j}^{[\ell]}$ being the update of the $\ell$th parameter of the $j$th coded symbol of the considered subpacket in $W_\theta$ and $\hat{z}_{\ell}$ is a random noise symbol. Once database $n$ receives all $U_{n,\ell}$, it calculates the incremental update as,
\begin{align}
    \bar{U}_{n,\ell}&=U_{n,\ell}\times \Tilde{D}_{n,\ell}\times Q_{n,\ell}\\
    &=\begin{bmatrix}
        \frac{1}{f_{1,\ell}-\alpha_n}\Delta_{\theta,1}^{[\ell]}e_M(\theta)+P_{\alpha_n}^{[1]}(y)\\
        \vdots\\
        \frac{1}{f_{y,\ell}-\alpha_n}\Delta_{\theta,y}^{[\ell]}e_M(\theta)+P_{\alpha_n}^{[y]}(y)
    \end{bmatrix}, \label{lemmm2}
\end{align}
where $\Tilde{D}_{n,\ell}$ is a scaling matrix given by,
\begin{align}           
    \Tilde{D}_{n,\ell}\!=\!\text{diag}\!\left(\frac{1_M}{\prod_{i=1}^K(f_{1,i}\!-\!\alpha_n)},\dotsc,\frac{1_M}{\prod_{i=1}^K(f_{y,i}\!-\!\alpha_n)}\right)
\end{align}
with the all ones vector of size $1\times M$ indicated as $1_M$.  Since the incremental updates ($\Bar{U}_{n,\ell}$) are of the same form as the storage in \eqref{storage}, the submodels are updated by $S_n^{[t]}=S_n^{[t-1]}+\sum_{\ell=1}^K \bar{U}_{n,\ell}$. The resulting writing and total costs are,
\begin{align}
    C_W&=\frac{K\times R}{K\times y}=\frac{2R}{R'-K-1},\\
    C_T&=\frac{2(R+R')}{R'-K-1}=\begin{cases}
        \frac{4R}{R-K-1}, & \text{if $R-K$ is odd}\\
        \frac{4R-2}{R-K-2}, & \text{if $R-K$ is even}.
    \end{cases}\label{total}
\end{align}

\subsection{Storage Mechanism}

The proposed storage mechanism consists of submodel partitioning and submodel encoding, where the parameters are encoded with a specific $(K,R)$ MDS code and stored at different subsets of databases in parts. In this section, we find the optimum coding parameters and fractions of submodels stored in each database to minimize the total cost.

\subsubsection{Submodel partitioning}

This determines the fractions of all $(K,R)$ MDS coded submodels to be stored in the $N$ databases with arbitrary storage constraints. Based on the $(K,R)$ MDS coded structure, each coded parameter must be stored in exactly $R$ databases. Let $\eta_i$ be the fraction of all submodels that are stored in the same subset of $R$ databases (the subset of databases indexed by $i$). Let $\mathcal{B}$ be the basis containing all $N\times1$ vectors with elements in $\{0,1\}$ with exactly $R$ ones, denoted by $\mathcal{B}=\{b_1,b_2,\dotsc,b_{|\mathcal{B}|}\}$.
Note that each $b_i$ corresponds to a specific subset of $R$ databases, indexed by $i$. Then, it is required to find the $\eta_i$s that satisfy,
\begin{align}
    \frac{1}{K}\sum_{i=1}^{|\mathcal{B}|}\eta_i b_i&=\mu, \quad \text{where} \quad  \mu=[\mu(1),\dotsc,\mu(N)]^T\label{p1}\\
    \sum_{i=1}^{|\mathcal{B}|}\eta_i&=1,\quad 
    0\leq\eta_1,\dotsc,\eta_{|\mathcal{B}|}\leq1, \label{p3}
\end{align}
to replicate each coded parameter at exactly $R$ databases, while ensuring that all databases are completely filled. The solution to this problem with uncoded parameters ($K=1$) is provided in \cite{heteroPIR} and \cite{utah} along with a necessary and sufficient condition for a solution to exist, given by,
\begin{align}\label{condition}
    \mu(n)\leq\frac{\sum_{n=1}^N\mu(n)}{R}, \quad n\in\{1,\dotsc,N\}.
\end{align}
In this work, we find the optimum coding parameters $K$ and $R$ that result in the minimum total cost of the PRUW process while satisfying \eqref{condition}, and solve \eqref{p1}-\eqref{p3} to find the partitions $\eta_i$ of coded submodels to be stored in each database. 

\subsubsection{Submodel encoding}\label{encode}

For a given set of storage constraints $\{\mu(n)\}_{n=1}^N$, the total available storage is $p=\sum_{n=1}^N \mu(n)$ which is essentially the maximum number of times each uncoded parameter in the model can be replicated in the system of databases. Let the parameters of the model be stored using a $(K,R)$ MDS code. Therefore, the total number of coded parameters in the model is $\frac{ML}{K}$, which allows each coded parameter to be replicated at a maximum of,
\begin{align}\label{repl}
    R\leq\frac{\sum_{n=1}^N\mu(n)ML}{ML/K}=Kp
\end{align}
databases, if $Kp\leq N$. At the same time, the storage in each database must satisfy,
\begin{align}
    \mu(n)ML\leq\frac{ML}{K}, \quad n\in\{1,\dotsc,N\} \ \implies \ K\leq\frac{1}{\Bar{\mu}}, 
\end{align}
where $\Bar{\mu}=\max_{n}\mu(n)$, to completely utilize the available space in each database. Therefore, the coding parameter $K$ must satisfy,
\begin{align}\label{condK}
    K\leq\min\left\{\frac{1}{\Bar{\mu}},\frac{N}{p}\right\}=\frac{1}{\Bar{\mu}},
\end{align}
as $p=\sum_{n=1}^N \mu(n)\leq\Bar{\mu}N$ implies $\frac{1}{\Bar{\mu}}\leq \frac{N}{p}$. Since the total cost of the PRUW scheme decreases with the number of replications $R$, see \eqref{total}, for given $K,p\in\mathbb{Z}^+$ satisfying \eqref{condK}, the optimum $R$ for the $(K,R)$ MDS code is given by $Kp$ from \eqref{repl}, and the resulting total cost is,
\begin{align}
    C_T(K,R)=\begin{cases}
    \frac{4Kp}{Kp-K-1}, \quad & \text{odd $K$, even $p$}\\
    \frac{4Kp-2}{Kp-K-2}, \quad & \text{otherwise},
    \end{cases}
\end{align}
which decreases with $K$. Therefore, for a given set of storage constraints $\{\mu(n)\}_{n=1}^N$, the minimum total cost is achieved by the $(K,R)$ MDS code in \eqref{storage} with $K$ and $R$ given by,
\begin{align}\label{opt}
    K=\frac{1}{\Bar{\mu}}=k, \quad  R=kp=r,
\end{align}
which automatically satisfies \eqref{condition} since
$\frac{\sum_{n=1}^N\mu(n)}{r}\!=\!\Bar{\mu}\!\geq\!\mu(n)$ for all $n$. However, since $k$ and $r$ are not necessarily integers, it requires additional calculations to obtain the optimum $(k,r)$ MDS codes with $k,r\in\mathbb{Z^+}$ that collectively result in the lowest total cost. Consider the general case where $k, r\notin\mathbb{Z}^+$. The general idea here is to divide each submodel into four sections and encode them using the four MDS codes given in Table~\ref{codes}. The sizes of the four sections are defined by the parameters $\alpha$, $\beta$ and $\delta$, where $0\leq \alpha,\beta,\delta\leq1$, as shown in Table~\ref{codes}. The total spaces allocated for the four MDS codes in the entire system of databases are given by,
\begin{align}
    \sum_{n=1}^N\hat{\mu}_1(n)&=\frac{\alpha\beta\lfloor r\rfloor}{\lfloor k\rfloor},\quad\quad
    \sum_{n=1}^N\hat{\mu}_2(n)\!=\!\frac{\alpha(1-\beta)\lceil r\rceil}{\lfloor k\rfloor}\label{eq1}\\
    \sum_{n=1}^N\!\bar{\mu}_1(n)&\!=\!\frac{(1\!-\!\alpha)\delta\lfloor r\rfloor}{\lceil k\rceil}, \quad \!\!\!\!\!
    \sum_{n=1}^N\!\bar{\mu}_2(n)\!=\!\frac{(1\!\!-\!\alpha)(1\!-\!\delta)\lceil r\rceil}{\lceil k\rceil}\label{eq4}
\end{align}

%\vspace{-0.35cm}
\begin{table}[t]
\begin{center}
\begin{tabular}{ |m{0.5cm}|m{2cm}|m{2cm}|m{2.3cm}| }
\hline
  case & MDS code & fraction of submodel & space allocated in database $n$\\ 
  \hline
  \text{1} & $(\lfloor k\rfloor, \lfloor r\rfloor ) $ & $\alpha\beta$ & $\hat{\mu}_1(n)$\\
  \hline
  \text{2} & $(\lfloor k\rfloor, \lceil r\rceil ) $ & $\alpha(1-\beta)$& $\hat{\mu}_2(n)$\\
  \hline
  \text{3} & $(\lceil k\rceil, \lfloor r\rfloor )$ & $(1-\alpha)\delta$& $\bar{\mu}_1(n)$\\
  \hline
  \text{4} & $(\lceil k\rceil, \lceil r\rceil )$  & $(1-\alpha)(1-\delta)$& $\bar{\mu}_2(n)$\\  
  \hline
\end{tabular}
\end{center}
\caption{Fractions of submodels and corresponding MDS codes.}
\label{codes}
\vspace*{-0.8cm}
\end{table}
%\vspace{-0.35cm}

The space allocated for each individual MDS code in each database must satisfy \eqref{condition} separately, to ensure that all databases are completely filled while also replicating each coded parameter at the respective number of databases, i.e.,
\begin{align}
    \hat{\mu}_1(n)&\leq\frac{\alpha\beta}{\lfloor k\rfloor},\qquad\qquad
    \hat{\mu}_2(n)\leq\frac{\alpha(1-\beta)}{\lfloor k\rfloor}\label{neq1}\\
    \bar{\mu}_1(n)&\leq\frac{(1-\alpha)\delta}{\lceil k\rceil},\qquad
    \bar{\mu}_2(n)\leq\frac{(1-\alpha)(1-\delta)}{\lceil k\rceil}.\label{neq4}
\end{align}
All four storage allocations in each database must satisfy,  
\begin{align}\label{sum}
\hat{\mu}_1(n)+\hat{\mu}_2(n)+\bar{\mu}_1(n)+\bar{\mu}_2(n)=\mu(n), \quad \forall n.
\end{align}
It remains to find the values of $\hat{\mu}_1(n)$, $\hat{\mu}_2(n)$, $\bar{\mu}_1(n)$, $\bar{\mu}_2(n)$, $\alpha$, $\beta$ and $\delta$ that minimize the total cost given by,
\begin{align}\label{min}
    \mathcal{C}&\!=\!\alpha\left(\beta C_T(\lfloor k\rfloor,\lfloor r\rfloor)+(1-\beta) C_T(\lfloor k\rfloor,\lceil r\rceil)\right)\nonumber\\
    &\quad\!+\!(1\!-\!\alpha)\left(\delta C_T(\lceil k\rceil,\lfloor r\rfloor)+(1\!-\!\delta) C_T(\lceil k\rceil,\lceil r\rceil)\right)
\end{align}
while satisfying \eqref{eq1}-\eqref{sum}, where $C_T$ is defined in \eqref{totalcost}. We consider two cases, 1) $\alpha=1$, 2) $\alpha<1$. When $\alpha=1$, only cases 1, 2 in Table~\ref{codes} are used in storage, and \eqref{condition} becomes $R\leq\lfloor k\rfloor p$, which results in $\lfloor k\rfloor p=s$ maximum replications. This replaces all ``$r$"s in Table~\ref{codes} and \eqref{eq1}-\eqref{neq4} by $s$ when $\alpha=1$.

\begin{lemma}\label{lem0}
When $\alpha=1$, $\beta$ is fixed at $\beta=\lceil s\rceil-s$ to satisfy \eqref{eq1} and \eqref{sum}, and $\hat{\mu}_1(n),\hat{\mu}_2(n)$ satisfying \eqref{neq1} are given by,
\begin{align}
    \hat{\mu}_1(n)&=\Tilde{m}(n)+(\mu(n)-\Tilde{m}(n)-\Tilde{h}(n))\Tilde{\gamma}\\
    \hat{\mu}_2(n)&=\Tilde{h}(n)+(\mu(n)-\Tilde{m}(n)-\Tilde{h}(n))(1-\Tilde{\gamma})
\end{align}
for all $n\in\{1,\dotsc,N\}$ where,
\begin{align}
    &\Tilde{m}(n)\!=\!\left[\mu(n)\!-\!\frac{s\!-\!\lfloor s\rfloor}{\lfloor k\rfloor}\right]^+\!\!\!\!,\quad \Tilde{h}(n)\!=\!\left[\mu(n)\!-\!\frac{\lceil s\rceil\!-\!s}{\lfloor k\rfloor}\right]^+\\
    &\Tilde{\gamma}=\frac{\frac{\lfloor s\rfloor}{\lfloor k\rfloor}(\lceil s\rceil-s)-\sum_{n=1}^N \Tilde{m}(n)}{p-\sum_{n=1}^N \Tilde{m}(n)-\sum_{n=1}^N \Tilde{h}(n)},
\end{align}
with $[x]^+\!=\!\max\{x,0\}$. The resulting total cost is given in \eqref{c1}.
\end{lemma}

\begin{lemma}\label{lem1}
The following values of $\hat{\mu}_1(n)$, $\hat{\mu}_2(n)$, $\bar{\mu}_1(n)$ and $\bar{\mu}_2(n)$ for $n\in\{1,\dotsc,N\}$ satisfy \eqref{eq1}-\eqref{sum} with any $\alpha<1$, $\beta$ and $\delta$ that satisfy $\alpha\geq\frac{\lfloor k\rfloor}{k}(\lceil k\rceil-k)$, $\beta\geq\left[1-\frac{\lfloor k\rfloor}{k\alpha}(r-\lfloor r\rfloor)\right]^+$ and $\delta\geq\left[1-\frac{\lceil k\rceil}{k(1-\alpha)}(r-\lfloor r\rfloor)\right]^+$.
\begin{align}\label{e1}
    \hat{\mu}_1(n)&\!=\!\begin{cases}\!
    \hat{\mu}(n)\beta,\!\! & \!\!\text{if $\beta\!\in\!\{0,1\}$}\\
    \!\hat{m}(n)\!+\!(\hat{\mu}(n)\!-\!\hat{m}(n)\!-\!\hat{h}(n))\hat{\gamma}, \!\!&\!\! \text{if $\beta\!\in\!(0,1)$}
    \end{cases}\\
    \hat{\mu}_2(n)&\!=\!\begin{cases}
    \hat{\mu}(n)(1-\beta), & \text{if $\beta\in\{0,1\}$}\\
    \hat{h}(n)\!+\!(\hat{\mu}(n)\!-\!\hat{m}(n)\!-\!\hat{h}(n))(1\!-\!\hat{\gamma}), & \text{if $\beta\in(0,1)$}
    \end{cases}\\
    \bar{\mu}_1(n)&\!=\!\begin{cases}
    \!\bar{\mu}(n)\delta, \!\!&\!\! \text{if $\delta\!\in\!\{0,1\}$}\\
    \!\bar{m}(n)\!+\!(\bar{\mu}(n)\!-\!\bar{m}(n)\!-\!\bar{h}(n))\bar{\gamma}, \!\!&\!\! \text{if $\delta\!\in\!(0,1)$}
    \end{cases}\\
    \bar{\mu}_2(n)&=\!\begin{cases}
    \bar{\mu}(n)(1\!-\!\delta), &\!\!\! \text{if $\delta\in\{0,1\}$}\\
    \bar{h}(n)\!+\!(\bar{\mu}(n)\!-\!\bar{m}(n)\!-\!\bar{h}(n))(1\!-\!\bar{\gamma}), & \text{if $\delta\in(0,1)$}
    \end{cases}\label{e2}   
\end{align}
for all $n\in\{1,\dotsc,N\}$, where,
\begin{align}
    \hat{\mu}(n)&=m(n)+(\mu(n)-m(n)-h(n))\gamma\label{ee1}\\
    \bar{\mu}(n)&=h(n)+(\mu(n)-m(n)-h(n))(1-\gamma)\\
    m(n)&\!=\!\left[\mu(n)\!-\!\frac{1-\alpha}{\lceil k\rceil}\right]^+,\quad h(n)\!=\!\left[\mu(n)\!-\!\frac{\alpha}{\lfloor k\rfloor}\right]^+\\    
    \gamma&=\frac{\frac{\alpha}{\lfloor k\rfloor}(\lceil r\rceil-\beta)-\sum_{n=1}^N m(n)}{p-\sum_{n=1}^N m(n)-\sum_{n=1}^N h(n)}\\
    \hat{m}(n)&\!=\!\left[\hat{\mu}(n)\!-\!\frac{\alpha(1-\beta)}{\lfloor k\rfloor}\right]^+\!\!,\quad\!\! \hat{h}(n)\!=\!\left[\hat{\mu}(n)\!-\!\frac{\alpha\beta}{\lfloor k\rfloor}\right]^+\\
    \hat{\gamma}&=\frac{\frac{\alpha\beta}{\lfloor k\rfloor}\lfloor r\rfloor-\sum_{n=1}^N \hat{m}(n)}{\frac{\alpha}{\lfloor k\rfloor}(\lceil r\rceil-\beta)-\sum_{n=1}^N \hat{m}(n)-\sum_{n=1}^N \hat{h}(n)}\\
    \bar{m}(n)&\!=\!\left[\bar{\mu}(n)\!-\!\frac{(1\!-\!\alpha)(1\!-\!\delta)}{\lceil k\rceil}\right]^+\!\!\!\!,\quad\!\! \bar{h}(n)\!=\!\left[\bar{\mu}(n)\!-\!\frac{(1\!-\!\alpha)\delta}{\lceil k\rceil}\right]^+\\
    \bar{\gamma}&=\frac{\frac{(1-\alpha)\delta}{\lceil k\rceil}\lfloor r\rfloor-\sum_{n=1}^N \bar{m}(n)}{\frac{1-\alpha}{\lceil k\rceil}(\lceil r\rceil-\delta)-\sum_{n=1}^N \bar{m}(n)-\sum_{n=1}^N \bar{h}(n)}.\label{ee2}
\end{align}
\end{lemma}

\begin{lemma}\label{lem2}
For the case where $\alpha<1$, the values of $\alpha$, $\beta$ and $\delta$ that minimize the total cost in \eqref{min} while satisfying the constraints in Lemma~\ref{lem1} are specified in Theorem~\ref{thm1}, and the corresponding total cost is given in \eqref{c2}.
\end{lemma}

Once all parameters in Table~\ref{codes} are determined from Lemmas~\ref{lem0}-\ref{lem2}, the methods proposed in \cite{heteroPIR} and \cite{utah} are used to find the solutions to \eqref{p1}-\eqref{p3}, i.e., $\eta_i$s for the four sets of storage allocations $\{\hat{\mu}_1\}_{n=1}^N$, $\{\hat{\mu}_2\}_{n=1}^N$, $\{\bar{\mu}_1\}_{n=1}^N$, $\{\bar{\mu}_2\}_{n=1}^N$ separately. These four sets of solutions determine where each individual coded symbol is replicated in the system of databases, which completes the storage mechanism.

As an example, consider a PRUW setting with $N=12$ databases, where $\mu(1)=\dotsc=\mu(5)=0.37$ and $\mu(6)=\dotsc=\mu(12)=0.35$. Note that the values of $k$, $r$ and $s$ here are the same as what is considered in Remark~\ref{rem1}, and therefore, result in the same $\alpha$, $\beta$, $\delta$ and total cost stated in Remark~\ref{rem1}. Using \eqref{e1}-\eqref{ee2}, we find the storage allocations as,
\begin{align}
    \hat{\mu}_1(n)&=\hat{\mu}(n)=\begin{cases}
        0.1107, & \text{for $n=1,\dotsc5$}\\
        0.0951, & \text{for $n=6,\dotsc12$}
    \end{cases}\label{dif1}\\
    \Bar{\mu}_1(n)&=\begin{cases}
        0.033, & \text{for $n=1,\dotsc5$}\\
        0.029, & \text{for $n=6,\dotsc12$}
    \end{cases}\label{dif2}\\
    \hat{\mu}_2(n)&=0,\quad \Bar{\mu}_2(n)=0.226,\quad \forall n.\label{same}
\end{align}

Note from \eqref{same} that the $(\lfloor k\rfloor, \lceil r\rceil)$ MDS code is not used, and the allocated space for the  $(\lceil k\rceil, \lceil r\rceil)$ MDS code in all databases is the same. Therefore, the optimum storage mechanism for PRUW with homogeneous storage constraints presented in \cite{pruw} is used to place the $(\lceil k\rceil, \lceil r\rceil)$ coded parameters. The storage allocations for $(\lfloor k\rfloor, \lfloor r\rfloor)$ and $(\lceil k\rceil, \lfloor r\rfloor)$ codes have different storage allocations for different databases as shown by the two different sets of values in \eqref{dif1} and \eqref{dif2}. Therefore, for these two cases, we solve \eqref{p1}-\eqref{p3} separately, to find the fractions of coded parameters to be stored in each database. For example, for the $(\lfloor k\rfloor, \lfloor r\rfloor)$ MDS code, we find the solution to \eqref{p1}-\eqref{p3} with $N=12$, $K=\lfloor k\rfloor=2$, $R=\lfloor r\rfloor=11$ and $\mu=[\hat{\mu}_1(1):\hat{\mu}_1(12)]$ as $\eta_1=\dotsc=\eta_5=2.73\times10^{-4}$ and $\eta_6=\dotsc=\eta_{12}=3.15\times10^{-2}$, where the $b_i$ in \eqref{p1} corresponding to each $\eta_i$ is the all ones vector of size $12\times1$ with a zero at the $i$th position. This essentially means that the storage mechanism assigns an $\eta_i$ fraction of $(\lfloor k\rfloor,\lfloor r\rfloor)$ coded parameters to the subset of $\lfloor r \rfloor=11$ databases that contain all databases except the $i$th one.

\newpage
\bibliographystyle{unsrt}
\bibliography{references}

\begin{thebibliography}{10}

\bibitem{pruw_jpurnal}
S.~Vithana and S.~Ulukus.
\newblock Private read update write ({PRUW}) in federated submodel learning
  ({FSL}): Communication efficient schemes with and without sparsification.
\newblock Available online at arXiv:2209.04421.

\bibitem{dropout}
Z.~Jia and S.~A. Jafar.
\newblock ${X}$-secure ${T}$-private federated submodel learning with elastic
  dropout resilience.
\newblock {\em IEEE Transactions on Information theory}, 68(8):5418--5439,
  August 2022.

\bibitem{pruw}
S.~Vithana and S.~Ulukus.
\newblock Private read update write {(PRUW)} with storage constrained
  databases.
\newblock In {\em IEEE ISIT}, June 2022.

\bibitem{sparse}
S.~Vithana and S.~Ulukus.
\newblock Private federated submodel learning with sparsification.
\newblock In {\em IEEE ITW}, November 2022.

\bibitem{rd}
S.~Vithana and S.~Ulukus.
\newblock Rate distortion tradeoff in private read update write in federated
  submodel learning.
\newblock In {\em Asilomar Conference}, October 2022.

\bibitem{ourICC}
S.~Vithana and S.~Ulukus.
\newblock Efficient private federated submodel learning.
\newblock In {\em IEEE ICC}, May 2022.

\bibitem{segmentation}
S.~Vithana and S.~Ulukus.
\newblock Model segmentation for storage efficient private federated learning
  with top $r$ sparsification.
\newblock Available online at arXiv:2212.11947.

\bibitem{rate_privacy_storage}
S.~Vithana and S.~Ulukus.
\newblock Rate-privacy-storage tradeoff in federated learning with top $r$
  sparsification.
\newblock Available online at arXiv:2212.09704.

\bibitem{secureFSL}
C.~Niu, F.~Wu, S.~Tang, L.~Hua, R.~Jia, C.~Lv, Z.~Wu, and G.~Chen.
\newblock Secure federated submodel learning.
\newblock Available online at arXiv:1911.02254.

\bibitem{paper1}
M.~Kim and J.~Lee.
\newblock Information-theoretic privacy in federated submodel learning.
\newblock Available online at arXiv:2008.07656.

\bibitem{billion}
C.~Niu, F.~Wu, S.~Tang, L.~Hua, R.~Jia, C.~Lv, Z.~Wu, and G.~Chen.
\newblock Billion-scale federated learning on mobile clients: A submodel design
  with tunable privacy.
\newblock In {\em MobiCom}, April 2020.

\bibitem{recent}
S.~Ulukus, S.~Avestimehr, M.~Gastpar, S.~A. Jafar, R.~Tandon, and C.~Tian.
\newblock Private retrieval, computing and learning: Recent progress and future
  challenges.
\newblock {\em IEEE JSAC}, 40(3):729--748, March 2022.

\bibitem{original}
B.~Chor, E.~Kushilevitz, O.~Goldreich, and M.~Sudan.
\newblock Private information retrieval.
\newblock {\em Journal of the ACM}, 45(6):965--981, November 1998.

\bibitem{sun2017capacity}
H.~Sun and S.~A. Jafar.
\newblock The capacity of private information retrieval.
\newblock {\em IEEE Transactions on Infomation Theory}, 63(7):4075--4088, July
  2017.

\bibitem{chaoTian}
C.~Tian, H.~Sun, and J.~Chen.
\newblock Capacity-achieving private information retrieval codes with optimal
  message size and upload cost.
\newblock {\em IEEE Transactions on Information Theory}, 65(11):7613--7627,
  November 2019.

\bibitem{Kumar_PIRarbCoded}
S.~Kumar, H.-Y. Lin, E.~Rosnes, and A.~G. i~Amat.
\newblock Achieving maximum distance separable private information retrieval
  capacity with linear codes.
\newblock {\em IEEE Transactions on Information Theory}, 65(7):4243--4273, July
  2019.

\bibitem{SPIR}
H.~Sun and S.~A. Jafar.
\newblock The capacity of symmetric private information retrieval.
\newblock {\em IEEE Transactions on Information Theory}, 65(1):322--329,
  January 2019.

\bibitem{PSI}
Z.~Wang, K.~Banawan, and S.~Ulukus.
\newblock Private set intersection: A multi-message symmetric private
  information retrieval perspective.
\newblock {\em IEEE Transactions on Information Theory}, 68(3):2001--2019,
  March 2022.

\bibitem{MMPIR}
K.~Banawan and S.~Ulukus.
\newblock Multi-message private information retrieval: Capacity results and
  near-optimal schemes.
\newblock {\em IEEE Transactions on Information Theory}, 64(10):6842--6862,
  October 2018.

\bibitem{YamamotoPIR}
T.~Chan, S.~Ho, and H.~Yamamoto.
\newblock Private information retrieval for coded storage.
\newblock In {\em IEEE ISIT}, June 2015.

\bibitem{VardyConf2015}
A.~Fazeli, A.~Vardy, and E.~Yaakobi.
\newblock Codes for distributed {PIR} with low storage overhead.
\newblock In {\em IEEE ISIT}, June 2015.

\bibitem{coded}
K.~Banawan and S.~Ulukus.
\newblock The capacity of private information retrieval from coded databases.
\newblock {\em IEEE Transactions on Information Theory}, 64(3):1945--1956,
  March 2018.

\bibitem{RobustPIR_Razane}
R.~Tajeddine and S.~El Rouayheb.
\newblock Robust private information retrieval on coded data.
\newblock In {\em IEEE ISIT}, June 2017.

\bibitem{MultiroundPIR}
H.~Sun and S.~A. Jafar.
\newblock Multiround private information retrieval: Capacity and storage
  overhead.
\newblock {\em IEEE Transactions on Information Theory}, 64(8):5743--5754,
  August 2018.

\bibitem{arbmsgPIR}
H.~Sun and S.~A. Jafar.
\newblock Optimal download cost of private information retrieval for arbitrary
  message length.
\newblock {\em IEEE Transactions on Information Forensics and Security},
  12(12):2920--2932, December 2017.

\bibitem{WangSkoglund}
Q.~{Wang} and M.~{Skoglund}.
\newblock On pir and symmetric pir from colluding databases with adversaries
  and eavesdroppers.
\newblock {\em IEEE Transactions on Information Theory}, 65(5):3183--3197, May
  2019.

\bibitem{colluding}
H.~Sun and S.~A. Jafar.
\newblock The capacity of robust private information retrieval with colluding
  databases.
\newblock {\em IEEE Transactions on Information Theory}, 64(4):2361--2370,
  April 2018.

\bibitem{codedcolluded}
R.~Freij-Hollanti, O.~Gnilke, C.~Hollanti, and D.~Karpuk.
\newblock Private information retrieval from coded databases with colluding
  servers.
\newblock {\em SIAM Journal on Applied Algebra and Geometry}, 1(1):647--664,
  2017.

\bibitem{codedcolludedJafar}
H.~Sun and S.~A. Jafar.
\newblock Private information retrieval from {MDS} coded data with colluding
  servers: Settling a conjecture by {F}reij-{H}ollanti et al.
\newblock {\em IEEE Transactions on Information Theory}, 64(2):1000--1022,
  February 2018.

\bibitem{tandon2017capacity}
R.~Tandon.
\newblock The capacity of cache aided private information retrieval.
\newblock In {\em Allerton Conference}, October 2017.

\bibitem{wei2017fundamental}
Y.-P. Wei, K.~Banawan, and S.~Ulukus.
\newblock Fundamental limits of cache-aided private information retrieval with
  unknown and uncoded prefetching.
\newblock {\em IEEE Transactions on Information Theory}, 65(5):3215--3232, May
  2019.

\bibitem{leaky}
I.~Samy, M.~Attia, R.~Tandon, and L.~Lazos.
\newblock Asymmetric leaky private information retrieval.
\newblock {\em IEEE Transactions on Information Theory}, 67(8):5352--5369,
  August 2021.

\bibitem{ravi-leaky}
I.~{Samy}, R.~{Tandon}, and L.~{Lazos}.
\newblock On the capacity of leaky private information retrieval.
\newblock In {\em IEEE ISIT}, pages 1262--1266, July 2019.

\bibitem{sideinfo}
S.~Kadhe, B.~Garcia, A.~Heidarzadeh, S.~El Rouayheb, and A.~Sprintson.
\newblock Private information retrieval with side information.
\newblock {\em IEEE Transactions on Information Theory}, 66(4):2032--2043,
  April 2020.

\bibitem{securePIRcapacity}
Q.~Wang, H.~Sun, and M.~Skoglund.
\newblock The capacity of private information retrieval with eavesdroppers.
\newblock {\em IEEE Transactions on Information Theory}, 65(5):3198--3214, May
  2019.

\bibitem{evesdroppers}
Q.~Wang, H.~Sun, and M.~Skoglund.
\newblock The capacity of private information retrieval with eavesdroppers.
\newblock {\em IEEE Transactions on Information Theory}, 65(5):3198--3214, May
  2019.

\bibitem{byzantine}
K.~Banawan and S.~Ulukus.
\newblock The capacity of private information retrieval from {B}yzantine and
  colluding databases.
\newblock {\em IEEE Transactions on Information Theory}, 65(2):1206--1219,
  February 2019.

\bibitem{semanticPIR}
S.~Vithana, K.~Banawan, and S.~Ulukus.
\newblock Semantic private information retrieval.
\newblock {\em IEEE Transactions on Information Theory}, 68(4):2635--2652,
  April 2022.

\bibitem{singleDB}
S.~Li and M.~Gastpar.
\newblock Single-server multi-message private information retrieval with side
  information: the general cases.
\newblock In {\em IEEE ISIT}, June 2020.

\bibitem{XSTPIR}
Z.~Jia and S.~A. Jafar.
\newblock {$X$}-secure {$T$}-private information retrieval from {MDS} coded
  storage with {B}yzantine and unresponsive servers.
\newblock {\em IEEE Transactions on Information Theory}, 66(12):7427--7438,
  December 2020.

\bibitem{CSA}
Z.~Jia, H.~Sun, and S.~A. Jafar.
\newblock Cross subspace alignment and the asymptotic capacity of {$X$}-secure
  {$T$}-private information retrieval.
\newblock {\em IEEE Transactions on Information Theory}, 65(9):5783--5798,
  September 2019.

\bibitem{smallfields}
J.~{Xu} and Z.~{Zhang}.
\newblock Building capacity-achieving {PIR} schemes with optimal
  sub-packetization over small fields.
\newblock In {\em IEEE ISIT}, pages 1749--1753, June 2018.

\bibitem{SecureStorage}
H.~Yang, W.~Shin, and J.~Lee.
\newblock Private information retrieval for secure distributed storage systems.
\newblock {\em IEEE Transactions on Information Forensics and Security},
  13(12):2953--2964, December 2018.

\bibitem{MAC_PIR_ITW2018}
K.~Banawan and S.~Ulukus.
\newblock Private information retrieval from multiple access channels.
\newblock In {\em IEEE ITW}, November 2018.

\bibitem{heteroPIR}
K.~{Banawan}, B.~{Arasli}, Y.-P. {Wei}, and S.~{Ulukus}.
\newblock The capacity of private information retrieval from heterogeneous
  uncoded caching databases.
\newblock {\em IEEE Transactions on Information Theory}, 66(6):3407--3416, June
  2020.

\bibitem{utah}
N.~Woolsey, R.~Chen, and M.~Ji.
\newblock Uncoded placement with linear sub-messages for private information
  retrieval from storage constrained databases.
\newblock {\em IEEE Transactions on Commmunications}, 68(10):6039--6053,
  October 2020.

\bibitem{efficient_storage_ITW2019}
K.~Banawan, B.~Arasli, and S.~Ulukus.
\newblock Improved storage for efficient private information retrieval.
\newblock In {\em IEEE ITW}, August 2019.

\bibitem{PIR_decentralized}
Y.-P. Wei, B.~Arasli, K.~Banawan, and S.~Ulukus.
\newblock The capacity of private information retrieval from decentralized
  uncoded caching databases.
\newblock {\em Information}, 10, December 2019.

\bibitem{Tandon2018PIRFS}
R.~Tandon, M.~Abdul-Wahid, F.~Almoualem, and D.~Kumar.
\newblock {PIR} from storage constrained databases - coded caching meets {PIR}.
\newblock {\em IEEE ICC}, May 2018.

\bibitem{StorageConstrainedPIR}
M.~A. Attia, D.~Kumar, and R.~Tandon.
\newblock The capacity of private information retrieval from uncoded storage
  constrained databases.
\newblock {\em IEEE Transactions on Information Theory}, 66(11):6617--6634,
  November 2020.

\bibitem{StorageConstrainedPIR_Wei}
Y.-P. {Wei} and S.~{Ulukus}.
\newblock The capacity of private information retrieval with private side
  information under storage constraints.
\newblock {\em IEEE Transactions on Information Theory}, 66(4):2023--2031,
  April 2020.

\end{thebibliography}
\end{document}